# Transformación Min-Max para la medición del rendimiento deportivo mediante el PIR

(Transformación Min-Max para medir el rendimiento deportivo)

**Resumen**

El objetivo del presente trabajo es el de analizar la utilidad que tiene transformar la información para medir el rendimiento deportivo. Más concretamente, se busca la normalización de los datos para evitar la dependencia que un índice puede mostrar de algunas de las variables que lo forman. Este análisis se realiza dentro del ámbito del baloncesto debido a la tradición existente en este deporte en la recogida de datos, si bien es fácilmente adaptable a cualquier otro deporte. Como resultado se propone una modificación del *Performance Index Rating* (PIR) para medir el rendimiento deportivo de forma que se evite su dependencia de los puntos anotados. Los resultados obtenidos se ilustran a partir de las estadísticas de Larry Bird, Earvin Johnson, Michael Jordan y Kobe Bryant a lo largo de su carrera y pueden servir para optimizar el proceso de renovación/despido/contratación de jugadores en el diseño de una plantilla o de ayuda en la toma de decisiones para otorgar galardones individuales como el de mejor jugador de una temporada.

**Palabras clave:** medición, rendimiento, normalización, re-escalamiento.

# Min-Max transformation for the measurement of sports performance through PIR

(Min-Max transformation for the measurement of sports performance)

**Abstract**

The objective of this work is to analyze the usefulness to transform the information to measure sports performance. This analysis is carried out within the field of basketball due to the existing tradition in this sport in data collection, although it is easily adaptable to any other sport. As a result, a modification of the Performance Index Rating (PIR) is proposed to measure sports performance. The results obtained are illustrated from the statistics of Larry Bird, Earvin Johnson, Michael Jordan and Kobe Bryant throughout their careers and can serve to optimize the process of player renewal/firing/hiring in the design of a roster or to aid decision-making in awarding individual awards as best player of a season.

## 1. Introducción

La medición del rendimiento individual es interesante desde varios puntos de vista. Algunos ejemplos de su utilidad pueden ser el establecimiento de salarios, ayuda en la decisión de renovaciones/contrataciones/despidos o la otorgación de premios individuales como el de mejor jugador de la temporada en una determinada liga. Por tal motivo, tal y como indican García-Rubio et al (2019), "el estudio de los indicadores de rendimiento técnico-tácticos en competición es una línea de investigación consolidada dentro de las Ciencias del Deporte ya que aporta información relevante y con gran aplicación práctica a los entrenadores (McGarry, 2009)".

Esta afirmación adquiere especial énfasis cuando el deporte analizado es el baloncesto debido a la más que consolidada cultura existente en esta disciplina a la hora de recoger información durante el desarrollo de un partido. Esta tradición rápidamente hizo suya al fenómeno *Moneyball* (Lewis, 2004), lo cual se ha visto reflejado en el uso de una gran variedad de técnicas para analizar la información disponible (ver, por ejemplo, Pérez-Sánchez et al, 2019) y que el caso particular de la generación de índices para medir el rendimiento deportivo haya sido ampliamente estudiado en la literatura existente (ver, por ejemplo, Berri, 1999; Martínez, J.A., 2010c, 2010b, 2010c; Casals y Martínez, 2013; Berri et al, 2015; Salmerón-Gómez y Gómez-Haro, 2016, 2019; Blanco et al, 2018; Marmarinos et al, 2019; Salmerón, R., 2020).

Una característica común a (prácticamente) todos los índices existentes es su dependencia de la anotación. Es habitual que jugadores meramente anotadores se sitúen en los primeros puestos de los rankings creados (Salmerón-Gómez y Gómez-Haro, 2019), cuando la lógica indica que si el índice está enfocado para medir el rendimiento global del jugador se ha de obtener una visión completa del rendimiento y no sólo de uno de sus aspectos. Un ejemplo de índice en el que se presenta este déficit (ver Salmerón-Gómez y Gómez-Haro, 2019) es el *Performance Index Rating* (PIR), el cual es ampliamente usado en Europa, ya sea en ligas nacionales como la española o en la máxima competición europea a nivel de clubes (Euroliga).

Con el objetivo de mitigar esta situación, en el presente trabajo se proponen dos modificaciones de este índice poniendo de manifiesto, por un lado, la idoneidad de transformar cada variable que conforma el índice y, por otro, la necesidad de establecer ponderaciones para fijar la importancia de cada una de ellas.

En este caso, se propone transformar las variables mediante un re-escalamiento Min-Max, ya que entonces las variables toman valores en un mismo rango que las equipara. Al mismo tiempo, se está estableciendo cuál es el rendimiento mínimo y máximo (como se verá a lo largo del trabajo, estos conceptos no se deben confundir con rendimientos bajos y altos), lo cual permite realizar comparaciones desde un punto de vista objetivo.

El trabajo se estructura como sigue: en la sección 2 se presenta la transformación de datos conocida como re-escalamiento Min-Max. En la sección 3 se proponen dos modificaciones, basadas en la transformación anterior, del PIR para medir el rendimiento deportivo, las cuales son ilustradas empíricamente en la sección 4 a partir de los datos generados por Larry Bird, Earvin Johnson, Michael Jordan y Kobe Bryant en las temporadas que han jugado en su carrera en la *National Basketball Association* (NBA). En la sección 5 se discuten los objetivos que se pueden conseguir al aplicar estos índices y en la sección 6 se destacan las principales conclusiones del presente trabajo.

## 2. Metodología: re-escalamiento Min-Max

En Mazziota y Pareto (2017) se recomienda que en el diseño de cualquier índice compuesto se ha de definir el fenómeno a medir, seleccionar las variables a usar en la composición del índice, determinar si éstas se han de normalizar y establecer cómo se agregan. También indican la utilidad de validar el índice compuesto creado para así determinar que se trata de una medida estable.

De entre todos estos requisitos, el presente apartado se centra en la normalización de variables y, más concretamente, en el re-escalamiento Min-Max. Este tipo de normalización transforma cualquier conjunto de datos **x** al intervalo [0,1] de la siguiente manera:

$$y_i = (x_i - \min(\mathbf{x}))/(\max(\mathbf{x})-\min(\mathbf{x})),$$

donde $x_i$ denota al elemento i-ésimo de **x** y min(**x**), max(**x**) al mínimo y máximo elemento de **x**, respectivamente.

Esta transformación, en el ambiente deportivo, se puede interpretar como sigue: los valores iguales a cero y uno se interpretan como los momentos de mínimo y máximo rendimiento de un individuo concreto en la variable re-escalada. Mientas que cualquier punto intermedio se puede ver como el porcentaje en que se supera el mínimo rendimiento. Por tanto, el establecimiento del valor mínimo y máximo es clave para la interpretación de los resultados obtenidos.

*2.1 Ejemplo 1*

El principal inconveniente de esta transformación es su sensibilidad a la presencia de datos anómalos entre los datos considerados. Sin embargo, como se ilustra en el presente ejemplo, el efecto de datos anómalos sobre la transformación difiere dependiendo de si éstos afectan al mínimo o máximo valor.

A continuación se consideran dos conjuntos de datos, el primero con un dato anómalo en la parte superior y otra en la inferior, y se re-escalan teniendo en cuenta dichos datos y sin hacerlo. Los datos y resultados se muestran en la siguiente tabla:

| $x_1$ | $x_2$ | $y_1$ | $y_2$ | $x_1$ | $x_2$ | $y_1$ | $y_2$ |
|---|---|---|---|---|---|---|---|
| 15 | 15 | 1 | 1 | | 15 | | 1 |
| 5 | 14 | 0.166 | 0.916 | 5 | 14 | 1 | 0.5 |
| 4 | 13 | 0.0625 | 0.833 | 4 | 13 | 0.5 | 0 |
| 3 | 3 | 0 | 0 | 3 | | 0 | |
| **Media** | | 0.305 | 0.687 | Media | | 0.5 | 0.5 |

Se observa que la presencia de datos anómalos referentes al máximo suponen una disminución del valor medio (de 0.5 se pasa a 0.305), mientras que los referentes al mínimo suponen un aumento del valor medio (de 0.5 se pasa a 0.687). Esto se debe a que, en ambos casos, la presencia de datos anómalos suponen que el denominador sea grande. Sin embargo, si el dato anómalo es referente al máximo, los numeradores serán pequeños, lo cual supone un cociente bajo. Pero si el dato anómalo hace referencia al mínimo, los numeradores serán grandes, lo cual supone una compensación con el denominador e, incluso, que el cociente sea grande.

*2.2 Ejemplo 2*

Para ilustrar la utilidad de esta transformación se van a considerar los datos referentes a las temporadas jugadas en la NBA por Larry Bird (LB, 13 temporadas), Earvin Johnson (EJ, 13 temporadas), Michael Jordan (MJ, 15 temporadas) y Kobe Bryant (KB, 20 temporadas) tanto en temporada regular como en las eliminatorias por el título (*playoff*). Dichos datos han sido obtenidos del portal web Basketball-Reference ([www.basketball-reference.com](www.basketball-reference.com)).

A partir de la información recogida, se ha calculado el PIR. Esta medida se calcula a partir de la siguiente expresión:

PIR = Puntos Anotados + Rebotes + Asistencias + Balones Recuperados + Tapones Realizados + Faltas Recibidas – Tiros de Campo Fallados – Tiros Libres Fallados – Balones Perdidos – Tapones Recibidos – Faltas Realizadas.

Adviértase que la información referente a las faltas recibidas y tapones recibidos no está disponible en Basketball-Reference, por tal motivo se ha considerado que dichas variables son iguales a cero.

Una vez calculado el PIR para cada jugador se ha re-escalado el mismo considerando como valores mínimos y máximos el mínimo y máximo PIR de cada jugador, obteniéndose el $PIR_{INDIVIDUAL}$. También se ha re-escalado el PIR de cada jugador considerando como valores mínimos y máximos el mínimo y máximo PIR de todos los jugadores conjuntamente, obteniéndose el $PIR_{CONJUNTO}$. Los valores medios en cada caso se muestran en la siguiente tabla:

Tabla 1: PIR individual y colectivo para LB, EJ, MJ y KB en liga regular y playoff

| Fase | PIR | LB | EJ | MJ | KB |
| --- | --- | --- | --- | --- | --- |
| Liga Regular | Individual | 0.614 (0.533) | 0.658 (0.555) | 0.527 (0.527) | 0.673 (0.652) |
| | Conjunto | 0.745 (0.722) | 0.744 (0.722) | 0.714 (0.751) | 0.475 (0.423) |
| Play | Individual | 0.657 (0.657) | 0.637 (0.649) | 0.551 (0.551) | 0.683 (0.536) |
| Off | Conjunto | 0.732 (0.591) | 0.794 (0.718) | 0.847 (0.766) | 0.504 (0.355) |

Se observa que desde el punto de vista individual es KB quien presenta los mejores valores medios individuales tanto en temporada regular como en *playoff*, sin embargo, cuando se consideran los cuatro jugadores en su conjunto, es precisamente éste jugador el que presenta los valores más bajos en ambos casos. Además, los otros tres jugadores aumentan sus medias en todos los casos.

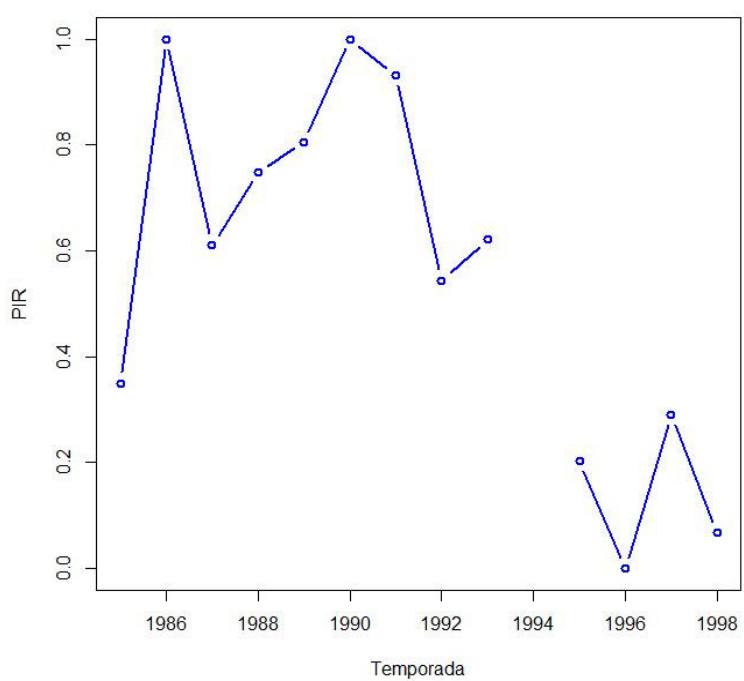

Figura 1: PIR$_{INDIVIDUAL}$ de MJ en playoffs

Considerando el caso de MJ en *playoff* se observa que tomando como referencia únicamente su rendimiento se obtiene la mínima media, mientras que si se considera como referencia el rendimiento del resto de jugadores, su rendimiento pasa a ser el mejor de todos. Se ponen de manifiesto entonces dos cuestiones:

- No tiene sentido usar los valores individuales para comparar el rendimiento entre jugadores. Su utilidad reside en obtener una trayectoria de rendimiento a lo largo de una temporada o vida deportiva. Así, por ejemplo, en la Figura 1 se representa el PIR$_{INDIVIDUAL}$ de MJ en los *playoffs*. Se observa que de los 6 campeonatos de la NBA que posee este jugador (1991, 1992, 1993, 1996, 1997 y 1998) sólo en el primero de ellos tiene un rendimiento excelente e, incluso, en el cuarto tiene su peor rendimiento en unos playoffs. Evidentemente, esto no significa que el rendimiento haya sido malo, sino el más bajo de todos en su trayectoria deportiva.

- Si el objetivo es el de comparar rendimientos entre jugadores, el significado de un bajo y alto rendimiento se ha de establecer de forma conjunta, para así poder contextualizar los resultados obtenidos. Así, se observa, que el rendimiento de MJ, comparado con el del resto de jugadores considerados, es el mejor de todos. Presentando una media del 84.7% superior al mínimo dato existente entre los cuatro jugadores.

Finalmente, destacar que entre paréntesis se presentan los valores medios sin considerar los siguientes datos anómalos:

- En liga regular: temporada 1988/89 de LB donde sólo juega 6 partidos; temporada 1995/96 de EJ que supone su vuelta tras su retirada por enfermedad; temporadas 1985/86, 2001/02 y 2002/03 de MJ donde, respectivamente, sólo juega 16 partidos y suponen su vuelta tras su segunda retirada; temporadas 1996/97 y 1997/98 de KB por presentar un rendimiento anormalmente bajo.

- En *playoff*: temporada 1995/96 de EJ y temporadas 1996/97 y 1997/98 de KB por idénticos motivos a los anteriores.

Se observa en general que los valores medios disminuyen debido a que (como se ha ilustrado en el ejemplo anterior) los datos anómalos afectan al mínimo, más concretamente, al bajo rendimiento de KB en sus dos primeras temporadas. Es decir, estos datos implican un beneficio más que una pérdida ya que suponen que los valores medios se inflen. En cualquier caso, no se producen cambios sustanciales en cuanto a la comparativa de jugadores realizada.

## 3. PIR re-escalado para medir el rendimiento

Como se ha comentado, un aspecto común a los índices generados para medir el rendimiento de un jugador o equipo en el baloncesto es su dependencia de los puntos anotados (Marmarinos et al, 2019; Salmerón-Gómez y Gómez-Haro, 2016, 2019). Para evitar dicha dependencia o mitigarla en cierto grado, en el presente trabajo se proponen dos índices.

El primero de ellos responde a la siguiente expresión:

$$PIR_{REES} = a_1*(\text{Puntos Anotados})_{REES} + a_2*\text{Rebotes}_{REES} + a_3*\text{Asistencias}_{REES} + a_4*(\text{Balones Recuperados})_{REES} + a_5*(\text{Tapones Realizados})_{REES} + a_6*(\text{Faltas Recibidas})_{REES} - a_7*(\text{Tiros de Campo Fallados})_{REES} - a_8*(\text{Tiros Libres Fallados})_{REES} - a_9*(\text{Balones Perdidos})_{REES} - a_{10}*(\text{Tapones Recibidos})_{REES} - a_{11}*(\text{Faltas Realizadas})_{REES},$$

donde el subíndice REES hace referencia a que dicha variable ha sido re-escalada (de forma individual o conjunta) según se ha mostrado en la sección anterior y los pesos $a_i$, i=1,…,11, son valores positivos que ponderan la relevancia de cada variable dentro del índice compuesto.

Además, este índice toma valores en el intervalo [$a_{MIN}$, $a_{MAX}$] donde $a_{MIN}$ = -($a_7 + a_8 + a_9 + a_{10} + a_{11}$) y $a_{MAX} = a_1 + a_2 + a_3 + a_4 + a_5 + a_6$. De forma que valores de $PIR_{REES}$ positivos indican que las acciones positivas superan a las negativas y viceversa para valores negativos de $PIR_{REES}$. Adviértase que si se considerase oportuno, $PIR_{REES}$ puede ser re-escalado para pasar de tomar valores en el intervalo [$a_{MIN}$, $a_{MAX}$] al intervalo [0, 1].

Por otro lado, en la construcción de este índice un paso fundamental es el de decidir los valores de las ponderaciones:

- En el caso $a_i = 1$ para todo i, se tiene que todas las variables consideradas en la elaboración del índice tienen la misma importancia.

- En el caso, por ejemplo, $a_1 = 3$, $a_2 = 2$, $a_i = 1$ para i=3,…,11, se tiene que los puntos anotados tienen un valor 1.5 veces superior a los rebotes y 3 veces superior que el resto de variables en la obtención de $PIR_{REES}$. De igual forma, los rebotes influyen el doble que el resto de variables (exceptuando los puntos anotados) en el cálculo de esta medida.

En definitiva se han de responder preguntas del tipo: ¿tiene el mismo valor un balón recuperado que un punto anotado? ¿y un rebote defensivo que otro defensivo? En caso negativo, ¿cuánto de más o de menos vale un balón recuperado que un punto anotado? ¿o un rebote defensivo que otro ofensivo? La respuesta no es evidente y seguramente difiera dependiendo del experto consultado.

Mientras que el segundo índice propuesto responde a la siguiente expresión:

$PIR_{POND}$ = (Puntos Anotados)$_{REES}$ * Puntos Anotados + Rebotes$_{REES}$ * Rebotes + Asistencias$_{REES}$ * Asistencias + (Balones Recuperados)$_{REES}$ * Balones Recuperados + (Tapones Realizados)$_{REES}$ * Tapones Realizados + (Faltas Recibidas)$_{REES}$ * Faltas Recibidas – (Tiros de Campo Fallados)$_{REES}$ * Tiros de Campo Fallados – (Tiros Libres Fallados)$_{REES}$ * Tiros Libres Fallados – (Balones Perdidos)$_{REES}$ * Balones Perdidos – (Tapones Recibidos)$_{REES}$ * Tapones Recibidos – (Faltas Realizadas)$_{REES}$ * Faltas Realizadas.

En este segundo caso se da respuesta a las cuestiones anteriores considerando que el valor de cada variable depende del entorno existente en cada momento y que queda recogido en el re-escalamiento de cada una de las variables. De esta forma se obtiene que esta ponderación cambia en cada temporada/partido ya que los puntos de referencia mínimo y máximo son distintos en cada caso.

En contrapartida se tiene que puede haber temporadas/partidos donde el peso de un punto anotado sea menor que el de un falta realizada, lo cual quizás tenga menos sentido que el hecho de que valgan lo mismo.

## 4. Resultados

A continuación se van a usar los datos referentes a LB, EJ, MJ y KB para mostrar las posibilidades del $PIR_{REES}$ y $PIR_{POND}$ en cuanto a la medición del rendimiento de un jugador.

*4.1 Ejemplo 3: todos los pesos en $PIR_{REES}$ son iguales*

En el caso de considerar que todos los pesos son iguales a uno se está considerando que todas las variables estadísticas tienen el mismo peso en el cálculo del índice propuesto para medir el rendimiento. Y puesto que recordemos que la información referente a las faltas y tapones recibidos no está disponible, los valores que puede tomar en este índice $PIR_{REES}$ son los del intervalo [-4, 5].

Los valores medios obtenidos en cada caso se muestran en la siguiente tabla:

Tabla 2: $PIR_{REES}$ individual y colectivo para LB, EJ, MJ y KB en liga regular y playoff

| Fase | $PIR_{REES}$ | LB | EJ | MJ | KB |
|------|--------------|----|----|----|----|
| Liga Regular | Individual | 1.4137 | 0.8914 | 1.215 | 1.357 |
| | Conjunto | 1.6264 | 1.304 | 0.822 | -0.0199 |
| Play Off | Individual | 0.942 | 0.809 | 1.164 | 1.2109 |
| | Conjunto | 1.179 | 0.943 | 0.451 | -0.1857 |

En el caso de que el rendimiento mínimo y máximo se establezca conjuntamente, se observa que al mitigar la dependencia del índice de los puntos anotados, el rendimiento de MJ disminuye sustancialmente con respecto a LB y EJ, siendo ambos los jugadores más completos de los cuatro considerados. Igualmente llaman la atención los valores medios negativos obtenidos por KB.

En la Figura 2 se representan los valores del $PIR_{REES}$ para LB en *playoff* tanto individual como conjunto. Se observa que los mejores valores en ambos casos se obtienen en las temporadas 1980/81 y 1985/86, temporadas en las que dicho jugador consiguió dos de sus tres campeonatos de la NBA.

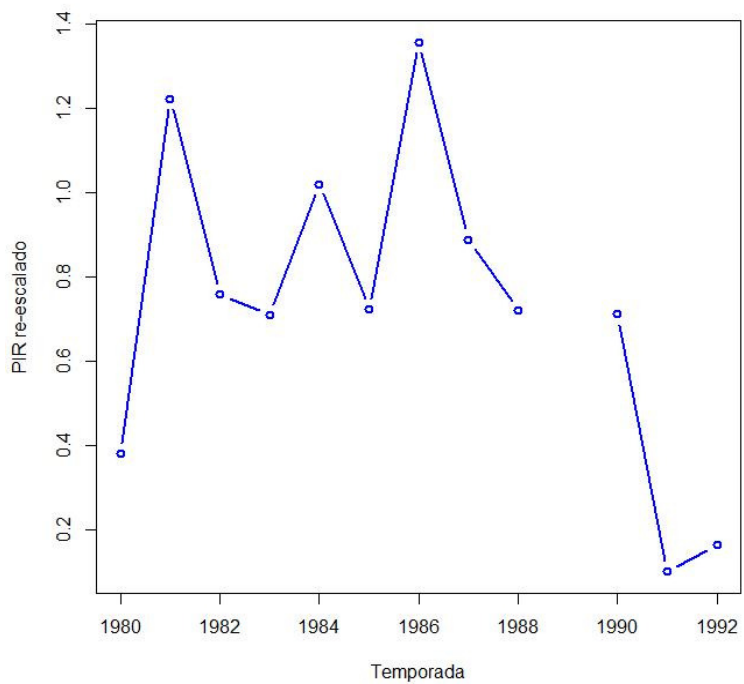

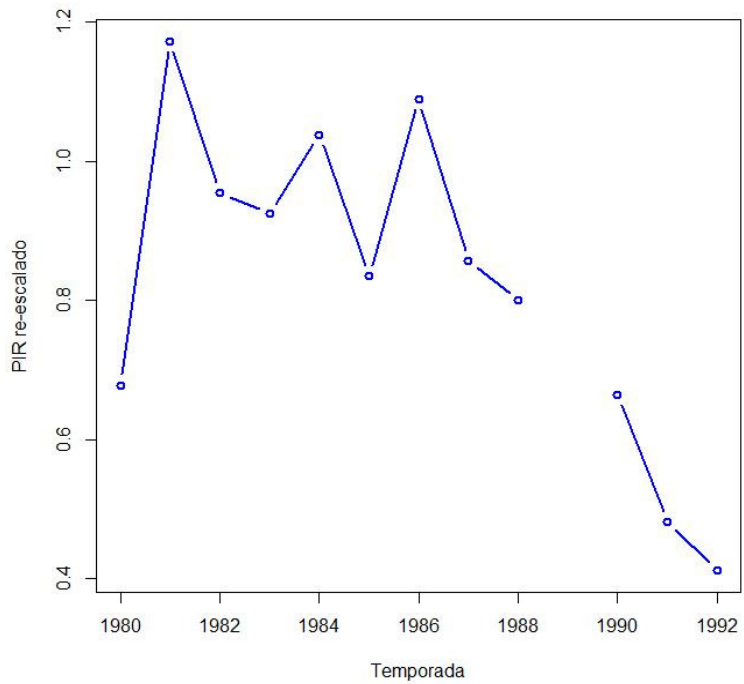

Figura 2: Valores de PIR$_{REES}$ para LB individual (arriba) y conjunto (abajo) en *playoff*

## 4.2 Ejemplo 4: los pesos en PIR$_{POND}$ son diferentes y determinados por el PIR$_{INDIVIDUAL}$

Si el re-escalado realizado en el ejemplo anterior se consideran como coeficientes en la expresión dada del PIR se obtiene el PIR$_{POND}$. Los valores medios obtenidos en cada caso se muestran en la Tabla 3.

Tabla 3: PIR$_{POND}$ individual y colectivo para LB, EJ, MJ y KB en liga regular y playoff

| Fase | PIR$_{POND}$ | LB | EJ | MJ | KB |
|---|---|---|---|---|---|
| Liga Regular | Individual | 15.995 | 16.213 | 17.035 | 13.569 |
|  | Conjunto | 17.627 | 17.281 | 17.515 | 8.662 |
| Play Off | Individual | 15.521 | 13.842 | 10.871 | 15.139 |
|  | Conjunto | 11.888 | 14.7633 | 16.719 | 5.656 |

Se observa un comportamiento similar al mostrado en la Tabla 1, donde en la liga regular hay un rendimiento muy similar entre LB, EJ y MJ, mientras que en *playoff* destaca el tercero de ellos una vez que se consideran los cuatro jugadores en su conjunto. Igualmente se observa que en la comparativa es KB el que presenta valores medios más bajos.

Comparando estos valores con los de la Tabla 2, se observa que MJ lidera de nuevo en el caso conjunto en los *playoffs*. Esto se debe a que los puntos anotados vuelven a tomar importancia en el cálculo del índice. En la Tabla 4 se muestran los valores mínimos y máximos para las anotaciones de cada jugador en *playoffs*. Se observa el dominio de MJ con respecto al resto de jugadores, incluso el mínimo valor de MJ es mayor que el máximo valor de LB y EJ.

Tabla 4: Anotaciones mínimas y máximas de cada jugador en *playoffs*

|  | LB | EJ | MJ | KB |
|---|---|---|---|---|
| Mínimo | 11.3 | 15.3 | 29.3 | 8.2 |
| Máximo | 27.5 | 25.2 | 43.7 | 32.8 |

Este dominio se traduce en que las ponderación media de los puntos anotados en $PIR_{POND}$ sea de 0.3915 para LB, 0.3113 para EJ, 0.7287 para MJ y 0.4702 para KB. Es decir, MJ recupera el protagonismo en el caso conjunto por su dominio anotador.

## 5. Discusión

Recientemente, en Salmerón (2020) se plantea la posibilidad de medir el rendimiento de un deportista a partir del rendimiento que se espera del mismo, de forma que se detecten comportamientos que aún siendo buenos se consideren insuficientes por estar por debajo de lo esperado. Y viceversa, comportamientos que siendo discretos se consideren como buenos por superar las expectativas planteadas.

Los índices presentados en el presente trabajo para medir el rendimiento de un jugador de baloncesto siguen la línea comentada en el sentido de que la calidad del rendimiento evaluado es determinado por el contexto que rodea al jugador. Es decir, no es lo mismo un jugador que anota 10 puntos por partido en una liga donde el 90% de los jugadores anotan al menos 6 puntos por partido o 16. En el primer caso se estaría entre el 10% de los jugadores que más puntos anotan mientras que en el segundo no.

En este sentido, se propone que el contexto sea establecido desde un punto de vista individual o conjunto:

- El primer caso resulta útil para analizar estados de forma de jugadores dentro de una misma temporada o carrera deportiva, de forma que se obtenga información útil para dar de baja o alta a jugadores en una competición (a día de hoy es habitual contar con plantillas de jugadores amplias que se prestan a este tipo de actuaciones) o para contratar/renovar/despedir jugadores cuando se conforman la configuración del equipo.

- El segundo es idóneo para comparar el rendimiento entre distintos jugadores, por lo que podría usarse, por ejemplo, para decidir un fichaje en el caso de existir duda entre jugadores que ocupan similar posición dentro del campo. También tendría utilidad a la hora de otorgar galardones individuales como el de mejor jugador de una liga.

En ambos casos, valores bajos de los índices indican que el rendimiento es bajo dentro del contexto considerado, que es distinto de decir que el rendimiento es bajo. Es decir, puede ser que un jugador tenga un buen comportamiento, pero dentro del contexto considerado sea de los peores. Así, por ejemplo, si el 10% de los mejores anotadores tienen una media de 16 puntos o más, aquellos jugadores con 16 puntos anotados de media formará parte de los peores dentro del grupo de los mejores anotadores.

Para lograr estas bondades en los índices propuestos se realiza una transformación en las variables consideradas consistente en un re-escalamiento Min-Max, de forma que el mínimo se identifica con el menor rendimiento y el máximo con el mayor. Si los mínimos y máximos se establecen a partir de los datos observados para cada jugador de forma individual, se estaría en el primer caso. Mientras que si se establecen a partir de los mínimos y máximos de los datos observados para todos los jugadores conjuntamente, se estaría en el segundo.

Hay que tener en cuenta que el re-escalamiento Min-Max es sensible a los datos anómalos, por lo que los índices basados en esta transformación también lo serán. Por tal motivo, se ha analizado qué efectos pueden tener este tipo de datos en la medición del rendimiento al mismo tiempo que se aconseja realizar un análisis previo (mediante técnicas de estadística descriptiva tradicionales como el gráfico de caja con bigotes) para detectar este tipo de datos y, en caso de que existan, eliminarlos del estudio.

Además, se proponen dos versiones del PIR, $PIR_{REES}$ y $PIR_{POND}$, basadas en su cálculo una vez normalizados las variables que integran este índice o ponderadas estas de forma conveniente. Se observa que sólo en el primer caso se mitigaría la dependencia que existe de los puntos anotados habitualmente en los distintos índices existentes para medir el rendimiento individual de un jugador de baloncesto.

Finalmente, las posibilidades de estas herramientas para medir el rendimiento de un jugador se han ilustrado a partir de las estadísticas generadas en liga regular y *playoff* por Larry Bird, Earvin Johnson, Michael Jordan y Kobe Bryant en todas las temporadas que han jugado en la NBA. Si bien no es uno de los objetivos del trabajo, se obtiene que Bryant es el que peor comportamiento presenta de los cuatro. Mientras que una vez mitigada la influencia de la anotación, Bird es el que presenta mejor rendimiento.

Como limitaciones del presente trabajo y, por tanto, futuras líneas de trabajo/mejora, está el establecimiento de ponderaciones en el segundo índice propuesto que puedan ser contraitutivos. Por ejemplo, que el peso de una falta sea mayor que el de un punto anotado. Una posibilidad para evitar esta situación puede ser la de establecer restricciones entre los coeficientes como que el coeficiente de ciertas variables tiene que ser siempre mayor o igual que las del resto. Con tal objetivo, se podría considerar la opinión de distintos expertos como pueden ser los managers o entrenadores de cada equipo.

## 6. Conclusiones y aplicaciones prácticas

En el presente trabajo se proponen dos modificaciones del PIR con el objetivo de establecer un sistema de ponderación para determinar la importancia dentro de dicho índice de las distintas variables que participan en su elaboración. Dentro de la propuesta realizada es clave la transformación de dichas variables mediante el re-escalamiento Min-Max, ya que permite establecer puntos de referencia mínimos y máximos que facilitan, por ejemplo, analizar el estado de forma de un jugador dentro de una temporada (o en toda su carrera) o la comparación de distintos jugadores. Por tanto, esta información podría ser usada por entrenadores dentro de una temporada para evaluar el "estado de forma" de cada uno de sus jugadores en cualquier momento de la temporada o por directores generales como ayuda a tomar decisiones como la renovación/baja de jugadores o la contratación de nuevo talento.

## Agradecimientos

## Referencias